\documentclass[aps,prd,twocolumn,superscriptaddress,nofootinbib,floatfix,noshowpacs]{revtex4-2}
\pdfoutput=1
\newif\ifcomment
\usepackage{amsmath,amssymb}
\usepackage{color,hyperref,url}
\usepackage{listings}
\usepackage{slashed}
\usepackage[pdftex]{graphicx}
\usepackage{epstopdf}
\usepackage{epsfig}
\usepackage{grffile}
\usepackage{relsize}
\graphicspath{{./img/}}
\usepackage{soul}
\usepackage{xcolor}

\newcommand{\beq}{\begin{equation}}
\newcommand{\eeq}{\end{equation}}
\newcommand{\ba}{\begin{array}}
\newcommand{\ea}{\end{array}}
\newcommand{\bea}{\begin{align}}
\newcommand{\eea}{\end{align}}
\newcommand{\bi}{\begin{itemize}}
\newcommand{\ei}{\end{itemize}}
\newcommand{\ben}{\begin{enumerate}}
\newcommand{\een}{\end{enumerate}}
\newcommand{\bc}{\begin{center}}
\newcommand{\ec}{\end{center}}
\newcommand{\bl}{\begin{flushleft}}
\newcommand{\el}{\end{flushleft}}
\newcommand{\br}{\begin{flushright}}
\newcommand{\er}{\end{flushright}}

\begin{document}


\title{Unveiling the Inner Structure of the Pion’s First Excited State}
\author{Xiaobin Wang}%
\email{wangxiaobin@mail.nankai.edu.cn}
\affiliation{School of Physics, Nankai University, Tianjin 300071, China}

\author{Lei Chang} \email{leichang@nankai.edu.cn}
\affiliation{School of Physics, Nankai University, Tianjin 300071, China}

\date{\today}

\begin{abstract}
By capturing the characteristics of the Bethe-Salpeter amplitude for the pion excitation state, we construct an algebraic model to provide the overall features of the pion’s first excitation state parton distribution amplitude and distribution function. We find that, at the hadronic scale, the distribution amplitude of the excited state exhibits nodes, while the distribution function is unimodal, with a peak at $x=1/2$ and distinct concave and convex fluctuations in the valence region. These findings provide new insights into the partonic structure of excited mesons and contribute significantly to our understanding of hadronic excitations.
\end{abstract}
%

\maketitle

\section{introduction}\label{sec:int}
Exploring the internal structure of hadrons, from both experimental~\cite{Anderle:2021wcy,Accardi:2012qut,Accardi:2023chb} and theoretical perspectives, is one of the key research areas in hadronic physics. 
The ultimate aim is to develop the 
tomographic images of quarks and gluons inside hadrons,
which will not only elucidate how gluons bind quarks to form hadrons but also provide a deeper understanding of the nature of the strong interaction.

Pseudoscalar mesons, due to their close connection with chiral symmetry, have garnered considerable attention~\cite{Roberts:2021nhw}. As the lightest pseudoscalar meson, the pion has been extensively studied, and the current consensus is that its distribution amplitude (DA) exhibits the largest broadening compared to the conformal limit form~\cite{Chang:2013pq,Stefanis:2014nla,Holligan:2023rex}. Although there remains some debate regarding the pion's parton distribution function (PDF), particularly its asymptotic behavior in the valence region~\cite{Alberg:2024svo}, further experimental efforts are expected to reduce these uncertainties. In addition to the pion, systems exhibiting flavor asymmetry, such as the kaon~\cite{Hua:2020gnw,Raya:2021zrz,Xu:2024nzp}, as well as heavy-flavor mesons like the $\eta_c$~\cite{Blossier:2024wyx}, have also been studied, providing a clearer understanding of the interplay between dynamical chiral symmetry breaking and the Higgs mechanism. Despite these advancements, there has been relatively little research on the internal structure of radial excitations of pseudoscalar mesons.

This paper aims to conduct a preliminary exploration of the structure of the pion’s radial excitations and investigate how they differ from the ground state. This is valuable for several reasons. First, the formation of the pion’s excited states is closely tied to the confinement mechanism. The excitation mass is much larger than the threshold for a pair of light quarks, yet these excitations do not decay into free quarks, offering a unique opportunity to probe confinement. Second, the excited states of the pion are expected to contribute to light-by-light scattering processes, which are important in the calculation of the anomalous magnetic moment of the muon $g-2$ ~\cite{Miramontes:2024fgo}. This makes the study of pion excitations relevant for both theoretical understanding and experimental investigations in particle physics.

In principle, within the framework of Continuum Schwinger Methods (CSMs), excited-state amplitudes in Euclidean space can be obtained by solving the Bethe-Salpeter equation, and subsequently, the light-front properties can be explored using previously developed techniques~\cite{Ding:2019lwe}. However, this approach faces significant numerical challenges. Moreover, while Lattice QCD has developed various algorithms to study hadron structure~\cite{Ji:2013dva,Radyushkin:2016hsy}, attempts to apply it to excited states remain scarce~\cite{Gao:2021hvs}. To overcome these challenges, we adopt an algebraic model for a preliminary study of the pion’s radial excitations. This model incorporates key features of the first radial excitation~\cite{Holl:2004fr,Qin:2011xq}, notably the presence of a node in the leading Chebyshev moment as the momentum evolves. Based on this model, we derive the distribution amplitude (DA), distribution function (DF), and transition form factor for the excited $\pi_{0}$ state , which form the main focus of our study.

The article is structured as follows: In Sec.\ref{sec:model}, we introduce the model of the quark propagator and the leading pseudoscalar Bethe-Salpeter amplitude. In Sec.\ref{sec:DA}, we discuss the distribution amplitude and present the $\gamma^{*}\pi\gamma$ transition form factor. In Sec.\ref{sec:DF}, we provide the distribution function and the charge density for the excited state. Finally, in Sec.\ref{sec:con}, we offer our conclusions and outlook.

\section{model}\label{sec:model}
The algebraic model we use here has been widely applied in the study of the internal structure of the pion ground state~\cite{Chang:2014lva,Chang:2016vkv,Xu:2018eii,Albino:2022gzs}. In summary, we use the simplest quark propagator,
\begin{equation}
S^{-1}(k)=i\gamma\cdot k+M\,,
\end{equation}
where $M$ represents the constituent mass. Additionally, we will use the Nakanishi representation~\cite{Nakanishi:1963zz} to model the meson's BS amplitude, and we only consider the leading term for the pseudoscalar meson,
\begin{equation}
\Gamma_{ps}(k;P)=i\gamma_{5}\frac{M^{2}}{\mathcal{N}}\int_{-1}^1 dz \rho(z)\frac{M^{2}-\beta(k+z P/2)^{2}}{[(k+z P/2)^{2}+M^{2}]^{2}}.
\end{equation}
Here, $\mathcal{N}=\dfrac{\sqrt{10(8-10\beta+7\beta^{2})}}{40\pi}$ the normalization constant to ensure the charge conservation and normalization of distribution function subsequently, $\beta$ is a parameter.

In the current algebraic model, considering that the excitation state mass is greater than twice the quark mass, singularities inevitably arise in the actual calculation. We tend to believe that mass effects are irrelevant to the current study. To simplify this difficulty, we have two provisional options: First, we can artificially increase the quark mass to more than half of the meson mass(like~\cite{Hobbs:2016xfz}); second, we can choose a fixed quark mass and set the meson mass to zero. The latter allows us to study both the ground state and the excited state simultaneously using the same quark propagator. And in this paper, we will adopt the second approach. The above-mentioned approach also implicitly reflects the fact that, due to confinement effects, mesons do not decay into a quark-antiquark pair.

The zeroth Chebyshev moment of the amplitude can be written
\begin{equation}
    \frac{2}{\pi}\int_0^{\pi}\sin^2\theta U_0(\cos\theta)\Gamma_{ps}(k;P)d\theta=i\gamma_{5}\frac{M^{2}(M^2-\beta k^2)}{\mathcal{N}(k^2+M^2)^2}\,,
\end{equation}
where $U_0(\cos\theta)=1$ is the zeroth order Chebyshev polynomial of the second kind and $k\cdot p=\cos\theta\sqrt{k^2p^2}$ (here we need to assume that $p^2\ne0$ firstly and make it zero in the final step). When $\beta=-1$, its form is consistent with the amplitude of the ground state. As $\beta$ deviates from $-1$ and its absolute value increases, it can characterize the features of the excitation state amplitude, i.e., the zeroth Chebyshev moment of the amplitude in momentum space exhibits a node~\cite{Holl:2004fr,Qin:2011xq}. In this paper, our goal is to hypothesize the general shape of the excitation state structure, so we choose the simplest spectral function~\cite{Chang:2013pq,Chang:2014lva},
\begin{equation}
\rho(z)=\dfrac{3}{4}(1-z^2)\,.
\end{equation}
It is noted that, for ground state in chiral limit~\cite{Chang:2014lva}, the $\text{DA}\sim x(1-x)$ and $\text{DF}\sim x^{2}(1-x)^{2}$, i.e., $\text{DF}\sim\text{DA}^{2}$. Consequently, within a unified framework, we can directly compare the differences between the DA and DF of the ground state and the excited state.

In order to determine parameter $\beta$ and $M$, let's first consider the decay constant of pseudoscalar meson $f_{ps}$, which is given by
\begin{equation}
f_{ps}P^{\mu}=\text{tr}_{\text{CD}}\int_{dq}^{\Lambda}\gamma_{5}\gamma^{\mu} \chi_{ps}(q-P/2;P),
\end{equation}
where, $P$ is the pion's total momentum; the trace is over color and spinor indices; $\int_{dq}$ is a Poincar\'e-invariant regularization of the four-dimensional integral. The BS wave function is
\begin{equation}
    \chi_{ps}(q-P/2;P)=S(q)\Gamma_{ps}(q-P/2;P)S(q-P),
\end{equation}
which can be calculated appropriately based on our model of quark propagator and BS amplitude.
After calculating, we obtain
\begin{equation}
    f_{ps}=\frac{(1-2\beta)M}{8\pi^{2}\mathcal{N}}\,.
\end{equation}

For the model parameters, we emphasize the following points:
\begin{itemize}
    \item $\beta=-1:$ ground state of pion. We can use the experimental value of the decay constant $92\text{MeV}$ to determine the model parameter $M=305\text{MeV}$. In the discussion on subsequent excited states, to maintain consistency, we will keep $M$ unchanged.
    \item $\beta=1/2:$ 1st excited state of pion in chiral limit. In this case the decay constant is zero~\cite{Holl:2004fr}.
    \item For the physical excited states, the experimental lower bound for the decay constant of the first excited state is given as $-5.9\text{MeV}$~\cite{ParticleDataGroup:2024cfk}. Using the above formula, we obtain $\beta=0.54$. Therefore, we will choose $\beta$ to lie within a certain range $(0.5,0.54)$ to determine the subsequent profiles for DA and DF.
\end{itemize}

\section{Distribution Amplitude and transition form factor}\label{sec:DA}

%
Considering that the decay constant of the excited state is zero in the chiral limit, and we wish to accommodate this fact~\cite{Holl:2004fr,Qin:2011xq}, we will define a DA with mass dimension, as follows~\cite{Li:2016dzv}
\begin{equation}
\varphi(x)=\text{tr}_{\text{CD}}\int_{dq}^{\Lambda}\delta(n\cdot q-x\ n\cdot P)\gamma_{5}\gamma\cdot n \chi_{ps}(q-P/2;P)
\end{equation}
where $n$ is a light-like vector ($n^2=0$). It is noted that the decay constant
\begin{equation}
f_{ps}=\int_{0}^{1}dx \varphi(x)\,.
\end{equation}

\begin{figure}[htbp]
    \centering
    \includegraphics[width=\linewidth]{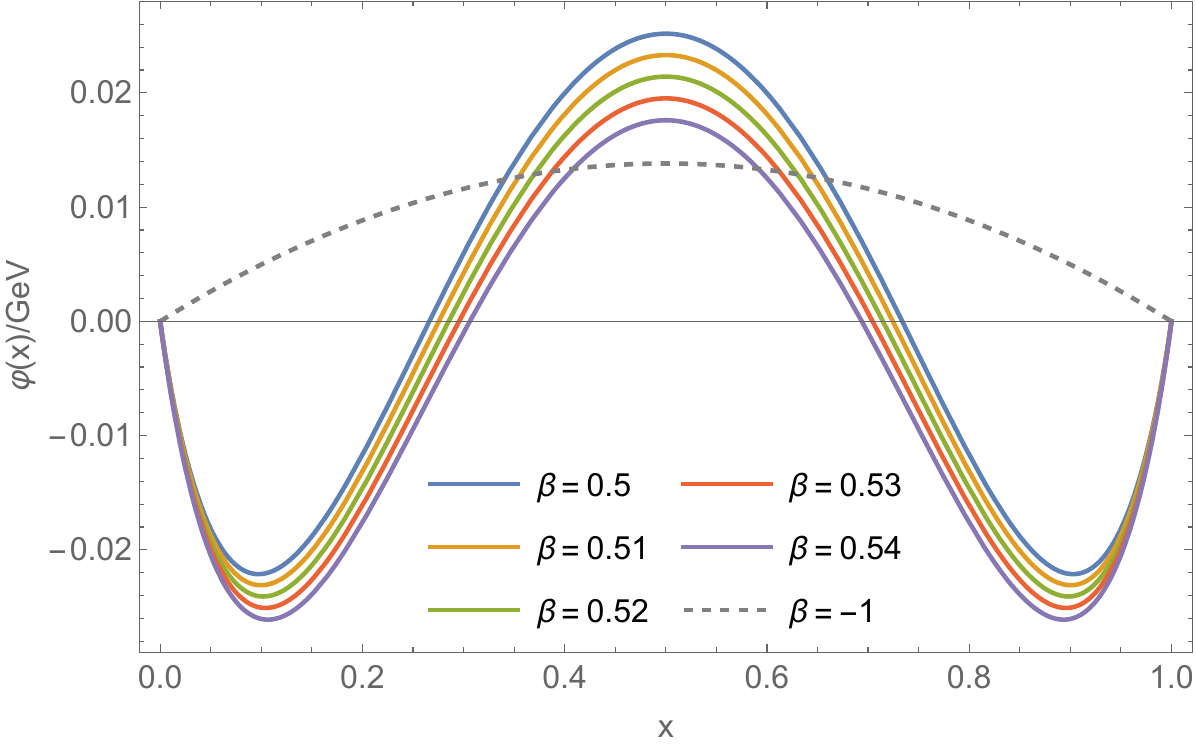}
    \caption{The distribution amplitude (DA) for the ground state and 1st excited state of the pion. The dashed line corresponds to the ground state with $\beta=-1$, and is divided by 10 for better comparison.}
    \label{DA}
\end{figure}

For such integrals, we need to use some tricks to calculate them in a roundabout manner. Let's first consider the moment of DA
\begin{equation}
\int_0^1x^m\varphi(x)dx=\text{tr}_{\text{CD}}\int_{dq}^{\Lambda}\frac{(n\cdot q)^m\gamma_{5}\gamma\cdot n \chi_{ps}(q-P/2;P)}{(n\cdot P)^{m+1}}\,,
\end{equation}
where $m$ is a nonnegative integer. Then using a Feynman parametrization to combine denominators, shifting the integration variable to isolate the integrations over Feynman parameters from that over the four-momentum $q$, and calculating the four-momentum integral, we obtain
\begin{equation}
    \begin{aligned}
        \int_0^1x^m\varphi(x)dx&=\int_{z,u_1,u_2}a^m\frac{3M[u_2(1+\beta)-\beta]\rho(z)}{4\pi^2\mathcal{N}}\\
        &=\int_{a,z,u_2}a^m\frac{9M[u_2(1+\beta)-\beta](1-z^2)}{16\pi^2\mathcal{N}},
    \end{aligned}
\end{equation}
where($a=u_1+(1-z)u_2/2$)
\begin{equation}
    \int_{z,u_1,u_2}=\int_{-1}^1dz\int_0^1du_1\int_0^{1-u_1}du_2,
\end{equation}
and
\begin{equation}
    \int_{a,z,u_2}=\int_0^1da\int_{-1}^1dz\int_0^{\text{Min}\{2x/(1-z),2(1-x)/(1+z)\}}du_2.
\end{equation}
By using the completeness of appropriate orthogonal polynomials (such as Legendre polynomials), it is not difficult to prove that if the moment sequences are equal, the integrand functions are equal. So finishing the integrations over $z$ and $u_2$, we can get the following expression of DA
\begin{equation}
    \begin{aligned}
        \varphi(x)=&-\frac{9M}{4\pi^{2}\mathcal{N}}\left[(1+2\beta)x(1-x)\right.\\
        &\left.+(1+\beta)(x^{2}\ln x+(1-x)^{2}\ln(1-x))\right].
    \end{aligned}
\end{equation}
According to the previously determined parameter range, we have drawn the image of DA as shown in Fig.~\ref{DA}.

Fig.~\ref{DA} illustrates the general properties of the 1st excited state DA~\cite{Li:2016dzv,Li:2021jqb}, which, in our model, is entirely attributed to the presence of node in the BS amplitude. The symmetry is natural because we have chosen isospin symmetry. Like the ground state, the maximum occurs at $x=1/2$, and as $x$ increases, the value changes from positive to negative. The position of the node is related to the size of beta (which suggests it is connected to the node position of the BS amplitude). As $x$ approaches $1$, DA tends to zero with a $(1-x)^1$ power law from below. This behavior arises partly because we have defined the decay constant as negative, and partly because the BS amplitude exhibits a $1/k^{2}$ behavior at large momentum.

It is well known that the DA of mesons is related to the ultraviolet behavior of the meson-$\gamma$ transition form factor~\cite{Hoferichter:2020lap}. We do not expect to fully capture the overall behavior of the form factor with such a simple model, but we can provide a general picture. To this end, we use a triangle diagram (we also suppose the bared vertex ansatz for both virtual and real photon quark couplings) to calculate the form factor, which is defined as follows,
\begin{equation}
\epsilon_{\mu\nu\rho\sigma}k_{1}^{\rho}k_{2}^{\sigma} \frac{G(Q^{2})}{4\pi^2}=-2\text{tr}_{D}\int_{dk}\gamma_{\mu}S(k_{\alpha})\gamma_{\nu}\chi(k_{\beta};-k_{1}-k_{2})\,,
\end{equation}
with $k_{1}^{2}=Q^{2}$ and $k_{2}^{2}=0$ for virtual and real photon momentum respectively, and $k_{\alpha}=k-k_{1}+k_{2}$, $k_{\beta}=k-k_{1}/2+k_{2}/2$.
When $Q^{2}=0$, we can obtain the following result,
\begin{equation}
    G(Q^2=0)=\frac{1-\beta}{6 \mathcal{N} M}\,,
\end{equation}
under the previous choice of $\beta$, this is positive for both ground and excited states. Noted that $G_{\pi_{1}}(0)/G_{\pi}(0)\in (0.53,0.57)$. One can apply the techniques from Ref.~\cite{Sultan:2024mva} to analyze the large  $Q^{2}$ behavior of form factor and we find
\begin{equation}
    \begin{aligned}
        \lim_{Q^2\to\infty}Q^{2}\frac{G(Q^2)}{4\pi^2} &\propto2\int dx\frac{\varphi(x)}{3x}\\
        &=\frac{M}{4\pi^2\mathcal{N}}\left[(1+\beta)\pi^2-3(3+4\beta)\right]\,.
    \end{aligned}
\end{equation}
This behavior corresponds to the Brodsky-Lepage limit, which we find applies not only to the ground state but also to the first excited state~\cite{Cao:2021ddi}. Furthermore, the limit is positive for the ground state but negative for the first excited state. The ratio of the limits between the excited and ground states lies within the range $(-0.15,-0.22)$. As $Q^2$ increases, we conclude that while the form factor of the ground state remains positive, the form factor of the excited state, like the BS amplitude and DA, existes a node. It is important to note that this discussion pertains to the case of a single virtual photon, which differs from the two-virtual-photon case considered in Ref.~\cite{Holl:2005vu}.

\section{Distribution Function and charge density}\label{sec:DF}
\begin{figure}[htbp!]
    \centering
    \includegraphics[width=\linewidth]{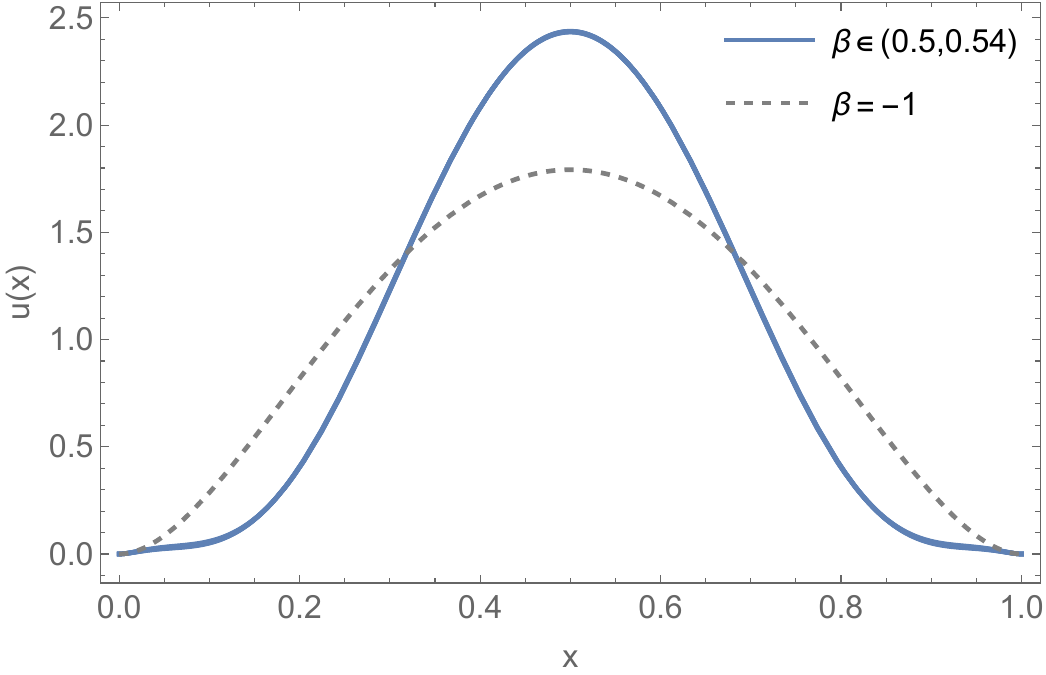}
    \includegraphics[width=\linewidth]{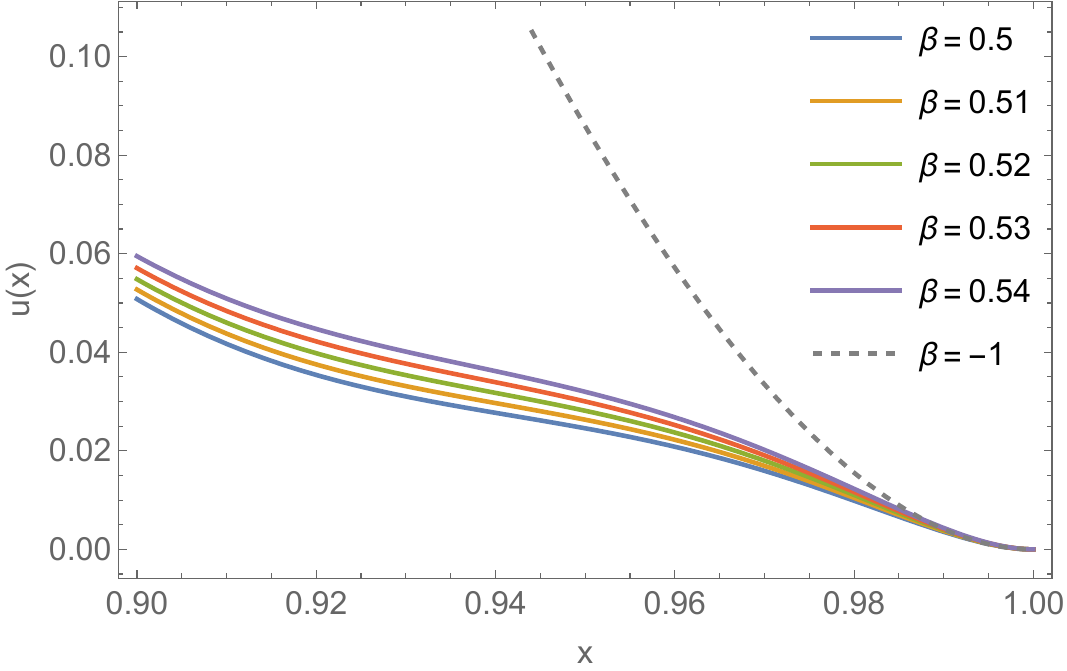}
    \caption{The distribution function (DF) for the ground state and 1st excited state of the pion. The DFs depicted in the upper pannel are indistinguishable in the chosen range of $\beta$.}
    \label{DF}
\end{figure}

We carry out the relevant calculations based on the definition of DF provided in Ref.~\cite{Chang:2014lva} for the case of momentum-dependent amplitudes. The advantage is that, at the hadronic scale, the fact that quarks carry all the hadron momentum can be ensured. Using the same techniques as in the previous section for calculating the DA, although somewhat more complex, the form of DF can be obtained analytically as follows,
\begin{widetext}
    \begin{equation}
        \begin{aligned}
            u(x)=&\frac{3(1+\beta)}{40\pi^2\mathcal{N}^2}\left\{-\frac{5\pi^2}{2}(15+19\beta)+\left[5\pi^2(33+41\beta)+\frac{327+830\beta+539\beta^2}{1+\beta}\right]x(1-x)\right.\\
            &-\left[20\pi^2(6+7\beta)+3\frac{329+770\beta+443\beta^2}{1+\beta}\right]x^2(1-x)^2\\
            &-15\left[15+19\beta-2(33+41\beta)x(1-x)+8(6+7\beta)x^2(1-x)^2\right]\ln x\ln(1-x)\\
            &-2\left[30(15+19\beta)-15(117+145\beta)x+5\frac{516+7\beta(158+83\beta)}{1+\beta}x^2+3\frac{128+3\beta(90+47\beta)}{1+\beta}(2x-5)x^3\right]x\ln x\\
            &+30\left[9+13\beta+6(1+\beta)x\right]x^4(\ln x)^2\\
            &\left.+30(2x-1)\left[15+19\beta-4(9+11\beta)x(1-x)+6(1+\beta)x^2(1-x)^2\right]\text{Li}_2(x)+(x\to1-x)\right\},
        \end{aligned}
    \end{equation}
\end{widetext}
where $\text{Li}_2(x)$ is the second order polylogarithm. When $\beta=-1$ the above expression matches the formula given in Ref.~\cite{Chang:2014lva}. Similarly, we have drawn the image of DF as shown in Fig.~\ref{DF}.

Fig.~\ref{DF} illustrates a lot of valuable properties of the 1st excited state DF. Firstly, like DA, DF is obviously symmetrical, both in terms of images and expressions, indicating the quarks carry all of the momentum of hadron. Compared with the ground state, the DF of the excited state is narrower and thinner, with a probability concentration towards 1/2. In the valence region, i.e., when $x$ approaches $1$, the excited state exhibits a fluctuation of concavity and convexity. Compared with DA, the excited state DF is very insensitive to changes in $\beta$. What's more, the excited state DF did not exhibit large-range oscillations like DA and disrupted the empirical rule that DF$\,\sim\,$DA$^2$, which works well for the ground state at hadronic scale. These two points have not been observed in previous studies.

\begin{figure}[htbp!]
    \centering
    \includegraphics[width=\linewidth]{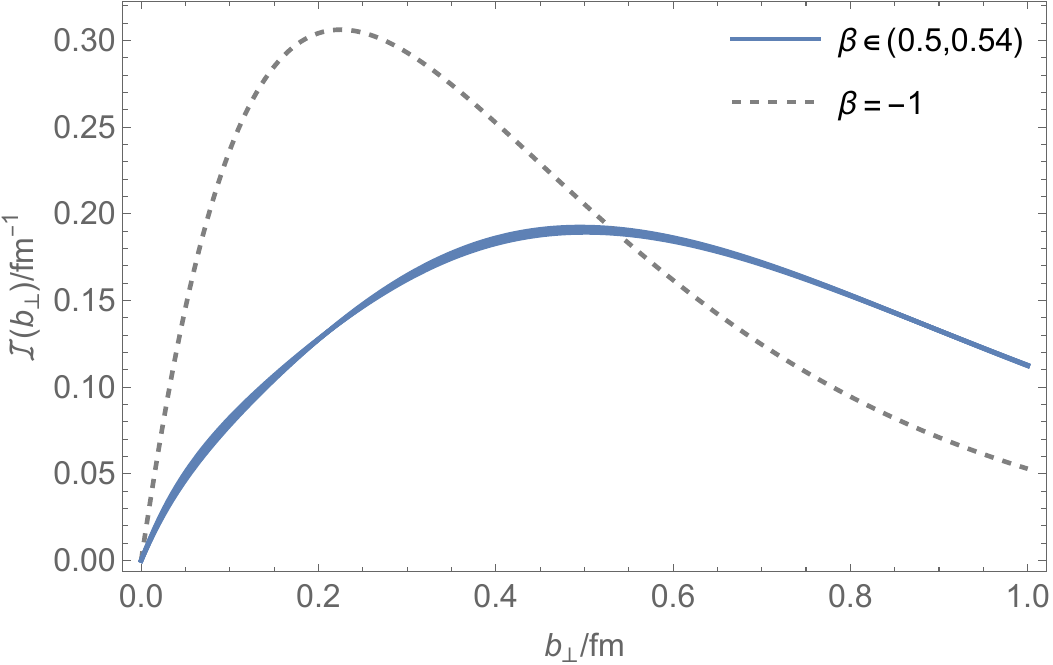}
    \caption{The charge density for the ground state and 1st excited state of the pion. Similar to DF, the charge densities are indistinguishable in the chosen range of $\beta$.}
    \label{CD}
\end{figure}

In order to a more intuitive depiction of the internal structure of the hadrons, let us now consider the charge density, $\rho(b_\perp)$, which can be obtained by performing the 2-dimensional Fourier transform of electromagnetic form factor $F(Q^2)$~\cite{Carmignotto:2014rqa},
\begin{equation}
    \begin{aligned}
        \rho(b_\perp)&=\int \frac{d^2Q}{(2\pi)^2}F(Q^2)e^{-i \bm{Q}\cdot \bm{b_\perp}}\\
        &=\int_0^\infty \frac{dQ}{2\pi}Q J_0(Q b_\perp) F(Q^2)\,,
    \end{aligned}
\end{equation}
where $J_0$ is the zeroth-order cylindrical Bessel function. For the electromagnetic form factor $F(Q^2)$, we use a triangle diagram to calculate it (the bared vertex ansatz is adopted for quark photon coupling again), which is definited as follow:
\begin{equation}
    \begin{aligned}
        K_{\mu}F(Q^2)=&\text{tr}_{CD}\int_{dq}\left\{i\Gamma_{ps}(q+p_f/2;-p_f)S(q+p_f)\right.\\
        &\left.\times i\gamma_{\mu}S(q+p_i)i\Gamma_{ps}(q+p_i/2;p_i)S(q)\right\}\,,
    \end{aligned}
\end{equation}
with $p_{i,f}^2=(K\mp Q/2)^2=0$ for the incoming and outgoing meson momentum respectively. The computed charge densities $\mathcal{I}(b_\perp)=2\pi b_\perp \rho(b_\perp)$ are shown in Fig.~\ref{CD}.

Fig.~\ref{CD} illustrates some valuable properties of the 1st excited state charge density. Firstly, the excited state charge density is very insensitive to changes in $\beta$, which is similar to DF. Compared with the ground state, the charge density distribution of the excited state is much broader, i.e., the excited state occupies a larger volume. This is in line with physical intuition; that is, the binding of excited states is weaker, and its structure is no longer tight. Quantitatively, the charge radius, which is definited as $\sqrt{-6F'(0)}$, changed from $0.736\,\text{fm}$ for the ground state to $1.106\,\text{fm}-1.112\,\text{fm}$ for the excited state, with a growth rate of $50.2\%-51\%$.

\section{Conclusion and outlook}\label{sec:con}
In this work, we have explored the internal structure of the first radial excitation of the pion, a pseudoscalar meson, using an algebraic model. Our study focused on the distribution amplitude (DA), distribution function (DF), charge density, and transition form factor of the excited state, highlighting the differences between these properties and those of the pion ground state. By employing a simplified model based on the continuum Schwinger methods and Nakanishi representation, we have gained initial insights into the nature of the pion's excited states.

The results reveal distinct features of the excited state, such as a node in the leading Chebyshev moment of the amplitude, which significantly alters the DA. A comparison between the DA of the ground state and the first excited state indicates that the excited state exhibits a broader DA with a more complex structure in momentum space. Furthermore, while the transition form factor was not fully captured due to the simplifications in the model, its asymptotic behavior was shown to be consistent with the Brodsky-Lepage limit, mirroring the behavior of the ground state. This suggests that similar dynamics govern the high-energy limit of both states, though the negative limit observed warrants further investigation within a more detailed framework.

The excited state’s DF, presented here for the first time, is narrower than the ground state DF, with a peak around $x=1/2$ and exhibits concave and convex fluctuations in the valence region. Unlike the DA, the DF of the excited state shows reduced sensitivity to changes in the parameter $\beta$ and does not follow the DF$\,\sim\,$DA$^2$ rule observed for the ground state at hadronic scale. These novel findings provide fresh insights into the structure of excited hadrons and add a new dimension to our understanding of their partonic distributions.

Although this preliminary study has shed light on the general features of excited-state mesons, several avenues for future research remain. First, the algebraic model presented here can be further refined by incorporating more complex quark-gluon interactions and utilizing advanced approaches such as lattice QCD simulations or Bethe-Salpeter equation calculations. These methods could offer a more precise and comprehensive understanding of the pion's excited states. Additionally, studying higher radial excitations, including their impact on processes like light-by-light scattering, offers intriguing possibilities for future theoretical and experimental investigations.

In conclusion, this work provides a foundation for future studies of excited pseudoscalar mesons, offering new insights into their internal structure and presenting a useful framework for understanding hadronic excitations in terms of their partonic distribution functions and form factors. Further theoretical advancements and experimental validation will be crucial to deepening our understanding of these fascinating aspects of hadronic physics.

\begin{acknowledgments}
This work is financially supported by the National Natural Science Foundation of China (grant no. 12135007).
\end{acknowledgments}

\bibliography{bibreferences}

\begin{thebibliography}{33}%
\makeatletter
\providecommand \@ifxundefined [1]{%
 \@ifx{#1\undefined}
}%
\providecommand \@ifnum [1]{%
 \ifnum #1\expandafter \@firstoftwo
 \else \expandafter \@secondoftwo
 \fi
}%
\providecommand \@ifx [1]{%
 \ifx #1\expandafter \@firstoftwo
 \else \expandafter \@secondoftwo
 \fi
}%
\providecommand \natexlab [1]{#1}%
\providecommand \enquote  [1]{``#1''}%
\providecommand \bibnamefont  [1]{#1}%
\providecommand \bibfnamefont [1]{#1}%
\providecommand \citenamefont [1]{#1}%
\providecommand \href@noop [0]{\@secondoftwo}%
\providecommand \href [0]{\begingroup \@sanitize@url \@href}%
\providecommand \@href[1]{\@@startlink{#1}\@@href}%
\providecommand \@@href[1]{\endgroup#1\@@endlink}%
\providecommand \@sanitize@url [0]{\catcode `\\12\catcode `\$12\catcode `\&12\catcode `\#12\catcode `\^12\catcode `\_12\catcode `\%12\relax}%
\providecommand \@@startlink[1]{}%
\providecommand \@@endlink[0]{}%
\providecommand \url  [0]{\begingroup\@sanitize@url \@url }%
\providecommand \@url [1]{\endgroup\@href {#1}{\urlprefix }}%
\providecommand \urlprefix  [0]{URL }%
\providecommand \Eprint [0]{\href }%
\providecommand \doibase [0]{https://doi.org/}%
\providecommand \selectlanguage [0]{\@gobble}%
\providecommand \bibinfo  [0]{\@secondoftwo}%
\providecommand \bibfield  [0]{\@secondoftwo}%
\providecommand \translation [1]{[#1]}%
\providecommand \BibitemOpen [0]{}%
\providecommand \bibitemStop [0]{}%
\providecommand \bibitemNoStop [0]{.\EOS\space}%
\providecommand \EOS [0]{\spacefactor3000\relax}%
\providecommand \BibitemShut  [1]{\csname bibitem#1\endcsname}%
\let\auto@bib@innerbib\@empty
\bibitem [{\citenamefont {Anderle}\ \emph {et~al.}(2021)\citenamefont {Anderle} \emph {et~al.}}]{Anderle:2021wcy}%
  \BibitemOpen
  \bibfield  {author} {\bibinfo {author} {\bibfnamefont {D.~P.}\ \bibnamefont {Anderle}} \emph {et~al.},\ }\bibfield  {title} {\bibinfo {title} {{Electron-ion collider in China}},\ }\href {https://doi.org/10.1007/s11467-021-1062-0} {\bibfield  {journal} {\bibinfo  {journal} {Front. Phys. (Beijing)}\ }\textbf {\bibinfo {volume} {16}},\ \bibinfo {pages} {64701} (\bibinfo {year} {2021})},\ \Eprint {https://arxiv.org/abs/2102.09222} {arXiv:2102.09222 [nucl-ex]} \BibitemShut {NoStop}%
\bibitem [{\citenamefont {Accardi}\ \emph {et~al.}(2016)\citenamefont {Accardi} \emph {et~al.}}]{Accardi:2012qut}%
  \BibitemOpen
  \bibfield  {author} {\bibinfo {author} {\bibfnamefont {A.}~\bibnamefont {Accardi}} \emph {et~al.},\ }\bibfield  {title} {\bibinfo {title} {{Electron Ion Collider: The Next QCD Frontier}: {Understanding the glue that binds us all}},\ }\href {https://doi.org/10.1140/epja/i2016-16268-9} {\bibfield  {journal} {\bibinfo  {journal} {Eur. Phys. J. A}\ }\textbf {\bibinfo {volume} {52}},\ \bibinfo {pages} {268} (\bibinfo {year} {2016})},\ \Eprint {https://arxiv.org/abs/1212.1701} {arXiv:1212.1701 [nucl-ex]} \BibitemShut {NoStop}%
\bibitem [{\citenamefont {Accardi}\ \emph {et~al.}(2024)\citenamefont {Accardi} \emph {et~al.}}]{Accardi:2023chb}%
  \BibitemOpen
  \bibfield  {author} {\bibinfo {author} {\bibfnamefont {A.}~\bibnamefont {Accardi}} \emph {et~al.},\ }\bibfield  {title} {\bibinfo {title} {{Strong interaction physics at the luminosity frontier with 22 GeV electrons at Jefferson Lab}},\ }\href {https://doi.org/10.1140/epja/s10050-024-01282-x} {\bibfield  {journal} {\bibinfo  {journal} {Eur. Phys. J. A}\ }\textbf {\bibinfo {volume} {60}},\ \bibinfo {pages} {173} (\bibinfo {year} {2024})},\ \Eprint {https://arxiv.org/abs/2306.09360} {arXiv:2306.09360 [nucl-ex]} \BibitemShut {NoStop}%
\bibitem [{\citenamefont {Roberts}\ \emph {et~al.}(2021)\citenamefont {Roberts}, \citenamefont {Richards}, \citenamefont {Horn},\ and\ \citenamefont {Chang}}]{Roberts:2021nhw}%
  \BibitemOpen
  \bibfield  {author} {\bibinfo {author} {\bibfnamefont {C.~D.}\ \bibnamefont {Roberts}}, \bibinfo {author} {\bibfnamefont {D.~G.}\ \bibnamefont {Richards}}, \bibinfo {author} {\bibfnamefont {T.}~\bibnamefont {Horn}},\ and\ \bibinfo {author} {\bibfnamefont {L.}~\bibnamefont {Chang}},\ }\bibfield  {title} {\bibinfo {title} {{Insights into the emergence of mass from studies of pion and kaon structure}},\ }\href {https://doi.org/10.1016/j.ppnp.2021.103883} {\bibfield  {journal} {\bibinfo  {journal} {Prog. Part. Nucl. Phys.}\ }\textbf {\bibinfo {volume} {120}},\ \bibinfo {pages} {103883} (\bibinfo {year} {2021})},\ \Eprint {https://arxiv.org/abs/2102.01765} {arXiv:2102.01765 [hep-ph]} \BibitemShut {NoStop}%
\bibitem [{\citenamefont {Chang}\ \emph {et~al.}(2013)\citenamefont {Chang}, \citenamefont {Cloet}, \citenamefont {Cobos-Martinez}, \citenamefont {Roberts}, \citenamefont {Schmidt},\ and\ \citenamefont {Tandy}}]{Chang:2013pq}%
  \BibitemOpen
  \bibfield  {author} {\bibinfo {author} {\bibfnamefont {L.}~\bibnamefont {Chang}}, \bibinfo {author} {\bibfnamefont {I.~C.}\ \bibnamefont {Cloet}}, \bibinfo {author} {\bibfnamefont {J.~J.}\ \bibnamefont {Cobos-Martinez}}, \bibinfo {author} {\bibfnamefont {C.~D.}\ \bibnamefont {Roberts}}, \bibinfo {author} {\bibfnamefont {S.~M.}\ \bibnamefont {Schmidt}},\ and\ \bibinfo {author} {\bibfnamefont {P.~C.}\ \bibnamefont {Tandy}},\ }\bibfield  {title} {\bibinfo {title} {{Imaging dynamical chiral symmetry breaking: pion wave function on the light front}},\ }\href {https://doi.org/10.1103/PhysRevLett.110.132001} {\bibfield  {journal} {\bibinfo  {journal} {Phys. Rev. Lett.}\ }\textbf {\bibinfo {volume} {110}},\ \bibinfo {pages} {132001} (\bibinfo {year} {2013})},\ \Eprint {https://arxiv.org/abs/1301.0324} {arXiv:1301.0324 [nucl-th]} \BibitemShut {NoStop}%
\bibitem [{\citenamefont {Stefanis}(2014)}]{Stefanis:2014nla}%
  \BibitemOpen
  \bibfield  {author} {\bibinfo {author} {\bibfnamefont {N.~G.}\ \bibnamefont {Stefanis}},\ }\bibfield  {title} {\bibinfo {title} {{What binds quarks together at different momentum scales? A conceptual scenario}},\ }\href {https://doi.org/10.1016/j.physletb.2014.10.018} {\bibfield  {journal} {\bibinfo  {journal} {Phys. Lett. B}\ }\textbf {\bibinfo {volume} {738}},\ \bibinfo {pages} {483} (\bibinfo {year} {2014})},\ \Eprint {https://arxiv.org/abs/1405.0959} {arXiv:1405.0959 [hep-ph]} \BibitemShut {NoStop}%
\bibitem [{\citenamefont {Holligan}\ \emph {et~al.}(2023)\citenamefont {Holligan}, \citenamefont {Ji}, \citenamefont {Lin}, \citenamefont {Su},\ and\ \citenamefont {Zhang}}]{Holligan:2023rex}%
  \BibitemOpen
  \bibfield  {author} {\bibinfo {author} {\bibfnamefont {J.}~\bibnamefont {Holligan}}, \bibinfo {author} {\bibfnamefont {X.}~\bibnamefont {Ji}}, \bibinfo {author} {\bibfnamefont {H.-W.}\ \bibnamefont {Lin}}, \bibinfo {author} {\bibfnamefont {Y.}~\bibnamefont {Su}},\ and\ \bibinfo {author} {\bibfnamefont {R.}~\bibnamefont {Zhang}},\ }\bibfield  {title} {\bibinfo {title} {{Precision control in lattice calculation of x-dependent pion distribution amplitude}},\ }\href {https://doi.org/10.1016/j.nuclphysb.2023.116282} {\bibfield  {journal} {\bibinfo  {journal} {Nucl. Phys. B}\ }\textbf {\bibinfo {volume} {993}},\ \bibinfo {pages} {116282} (\bibinfo {year} {2023})},\ \Eprint {https://arxiv.org/abs/2301.10372} {arXiv:2301.10372 [hep-lat]} \BibitemShut {NoStop}%
\bibitem [{\citenamefont {Alberg}\ and\ \citenamefont {Miller}(2024)}]{Alberg:2024svo}%
  \BibitemOpen
  \bibfield  {author} {\bibinfo {author} {\bibfnamefont {M.}~\bibnamefont {Alberg}}\ and\ \bibinfo {author} {\bibfnamefont {G.~A.}\ \bibnamefont {Miller}},\ }\bibfield  {title} {\bibinfo {title} {{Quark counting, Drell-Yan West, and the pion wave function}},\ }\href {https://doi.org/10.1103/PhysRevC.110.L042201} {\bibfield  {journal} {\bibinfo  {journal} {Phys. Rev. C}\ }\textbf {\bibinfo {volume} {110}},\ \bibinfo {pages} {L042201} (\bibinfo {year} {2024})},\ \Eprint {https://arxiv.org/abs/2403.03356} {arXiv:2403.03356 [hep-ph]} \BibitemShut {NoStop}%
\bibitem [{\citenamefont {Hua}\ \emph {et~al.}(2021)\citenamefont {Hua}, \citenamefont {Chu}, \citenamefont {Sun}, \citenamefont {Wang}, \citenamefont {Xu}, \citenamefont {Yang}, \citenamefont {Zhang},\ and\ \citenamefont {Zhang}}]{Hua:2020gnw}%
  \BibitemOpen
  \bibfield  {author} {\bibinfo {author} {\bibfnamefont {J.}~\bibnamefont {Hua}}, \bibinfo {author} {\bibfnamefont {M.-H.}\ \bibnamefont {Chu}}, \bibinfo {author} {\bibfnamefont {P.}~\bibnamefont {Sun}}, \bibinfo {author} {\bibfnamefont {W.}~\bibnamefont {Wang}}, \bibinfo {author} {\bibfnamefont {J.}~\bibnamefont {Xu}}, \bibinfo {author} {\bibfnamefont {Y.-B.}\ \bibnamefont {Yang}}, \bibinfo {author} {\bibfnamefont {J.-H.}\ \bibnamefont {Zhang}},\ and\ \bibinfo {author} {\bibfnamefont {Q.-A.}\ \bibnamefont {Zhang}} (\bibinfo {collaboration} {Lattice Parton}),\ }\bibfield  {title} {\bibinfo {title} {{Distribution Amplitudes of K* and \ensuremath{\phi} at the Physical Pion Mass from Lattice QCD}},\ }\href {https://doi.org/10.1103/PhysRevLett.127.062002} {\bibfield  {journal} {\bibinfo  {journal} {Phys. Rev. Lett.}\ }\textbf {\bibinfo {volume} {127}},\ \bibinfo {pages} {062002} (\bibinfo {year} {2021})},\ \Eprint {https://arxiv.org/abs/2011.09788} {arXiv:2011.09788 [hep-lat]} \BibitemShut {NoStop}%
\bibitem [{\citenamefont {Raya}\ \emph {et~al.}(2022)\citenamefont {Raya}, \citenamefont {Cui}, \citenamefont {Chang}, \citenamefont {Morgado}, \citenamefont {Roberts},\ and\ \citenamefont {Rodriguez-Quintero}}]{Raya:2021zrz}%
  \BibitemOpen
  \bibfield  {author} {\bibinfo {author} {\bibfnamefont {K.}~\bibnamefont {Raya}}, \bibinfo {author} {\bibfnamefont {Z.-F.}\ \bibnamefont {Cui}}, \bibinfo {author} {\bibfnamefont {L.}~\bibnamefont {Chang}}, \bibinfo {author} {\bibfnamefont {J.-M.}\ \bibnamefont {Morgado}}, \bibinfo {author} {\bibfnamefont {C.~D.}\ \bibnamefont {Roberts}},\ and\ \bibinfo {author} {\bibfnamefont {J.}~\bibnamefont {Rodriguez-Quintero}},\ }\bibfield  {title} {\bibinfo {title} {{Revealing pion and kaon structure via generalised parton distributions *}},\ }\href {https://doi.org/10.1088/1674-1137/ac3071} {\bibfield  {journal} {\bibinfo  {journal} {Chin. Phys. C}\ }\textbf {\bibinfo {volume} {46}},\ \bibinfo {pages} {013105} (\bibinfo {year} {2022})},\ \Eprint {https://arxiv.org/abs/2109.11686} {arXiv:2109.11686 [hep-ph]} \BibitemShut {NoStop}%
\bibitem [{\citenamefont {Xu}\ \emph {et~al.}(2024)\citenamefont {Xu}, \citenamefont {Binosi}, \citenamefont {Chen}, \citenamefont {Raya}, \citenamefont {Roberts},\ and\ \citenamefont {Rodr\'\i{}guez-Quintero}}]{Xu:2024nzp}%
  \BibitemOpen
  \bibfield  {author} {\bibinfo {author} {\bibfnamefont {Z.-N.}\ \bibnamefont {Xu}}, \bibinfo {author} {\bibfnamefont {D.}~\bibnamefont {Binosi}}, \bibinfo {author} {\bibfnamefont {C.}~\bibnamefont {Chen}}, \bibinfo {author} {\bibfnamefont {K.}~\bibnamefont {Raya}}, \bibinfo {author} {\bibfnamefont {C.~D.}\ \bibnamefont {Roberts}},\ and\ \bibinfo {author} {\bibfnamefont {J.}~\bibnamefont {Rodr\'\i{}guez-Quintero}},\ }\href@noop {} {\bibinfo {title} {{Kaon Distribution Functions from Empirical Information}}} (\bibinfo {year} {2024}),\ \Eprint {https://arxiv.org/abs/2411.15376} {arXiv:2411.15376 [hep-ph]} \BibitemShut {NoStop}%
\bibitem [{\citenamefont {Blossier}\ \emph {et~al.}(2024)\citenamefont {Blossier}, \citenamefont {Mangin-Brinet}, \citenamefont {Morgado~Ch\'avez},\ and\ \citenamefont {San~Jos\'e}}]{Blossier:2024wyx}%
  \BibitemOpen
  \bibfield  {author} {\bibinfo {author} {\bibfnamefont {B.}~\bibnamefont {Blossier}}, \bibinfo {author} {\bibfnamefont {M.}~\bibnamefont {Mangin-Brinet}}, \bibinfo {author} {\bibfnamefont {J.~M.}\ \bibnamefont {Morgado~Ch\'avez}},\ and\ \bibinfo {author} {\bibfnamefont {T.}~\bibnamefont {San~Jos\'e}},\ }\bibfield  {title} {\bibinfo {title} {{The distribution amplitude of the \ensuremath{\eta}$_{c}$-meson at leading twist from lattice QCD}},\ }\href {https://doi.org/10.1007/JHEP09(2024)079} {\bibfield  {journal} {\bibinfo  {journal} {JHEP}\ }\textbf {\bibinfo {volume} {09}},\ \bibinfo {pages} {079}},\ \Eprint {https://arxiv.org/abs/2406.04668} {arXiv:2406.04668 [hep-lat]} \BibitemShut {NoStop}%
\bibitem [{\citenamefont {Miramontes}\ \emph {et~al.}(2024)\citenamefont {Miramontes}, \citenamefont {Raya}, \citenamefont {Bashir}, \citenamefont {Roig},\ and\ \citenamefont {Paredes-Torres}}]{Miramontes:2024fgo}%
  \BibitemOpen
  \bibfield  {author} {\bibinfo {author} {\bibfnamefont {A.~S.}\ \bibnamefont {Miramontes}}, \bibinfo {author} {\bibfnamefont {K.}~\bibnamefont {Raya}}, \bibinfo {author} {\bibfnamefont {A.}~\bibnamefont {Bashir}}, \bibinfo {author} {\bibfnamefont {P.}~\bibnamefont {Roig}},\ and\ \bibinfo {author} {\bibfnamefont {G.}~\bibnamefont {Paredes-Torres}},\ }\href@noop {} {\bibinfo {title} {{Radially excited pion: electromagnetic form factor and the box contribution to the muon's $g-2$}}} (\bibinfo {year} {2024}),\ \Eprint {https://arxiv.org/abs/2411.02218} {arXiv:2411.02218 [hep-ph]} \BibitemShut {NoStop}%
\bibitem [{\citenamefont {Ding}\ \emph {et~al.}(2020)\citenamefont {Ding}, \citenamefont {Raya}, \citenamefont {Binosi}, \citenamefont {Chang}, \citenamefont {Roberts},\ and\ \citenamefont {Schmidt}}]{Ding:2019lwe}%
  \BibitemOpen
  \bibfield  {author} {\bibinfo {author} {\bibfnamefont {M.}~\bibnamefont {Ding}}, \bibinfo {author} {\bibfnamefont {K.}~\bibnamefont {Raya}}, \bibinfo {author} {\bibfnamefont {D.}~\bibnamefont {Binosi}}, \bibinfo {author} {\bibfnamefont {L.}~\bibnamefont {Chang}}, \bibinfo {author} {\bibfnamefont {C.~D.}\ \bibnamefont {Roberts}},\ and\ \bibinfo {author} {\bibfnamefont {S.~M.}\ \bibnamefont {Schmidt}},\ }\bibfield  {title} {\bibinfo {title} {{Symmetry, symmetry breaking, and pion parton distributions}},\ }\href {https://doi.org/10.1103/PhysRevD.101.054014} {\bibfield  {journal} {\bibinfo  {journal} {Phys. Rev. D}\ }\textbf {\bibinfo {volume} {101}},\ \bibinfo {pages} {054014} (\bibinfo {year} {2020})},\ \Eprint {https://arxiv.org/abs/1905.05208} {arXiv:1905.05208 [nucl-th]} \BibitemShut {NoStop}%
\bibitem [{\citenamefont {Ji}(2013)}]{Ji:2013dva}%
  \BibitemOpen
  \bibfield  {author} {\bibinfo {author} {\bibfnamefont {X.}~\bibnamefont {Ji}},\ }\bibfield  {title} {\bibinfo {title} {{Parton Physics on a Euclidean Lattice}},\ }\href {https://doi.org/10.1103/PhysRevLett.110.262002} {\bibfield  {journal} {\bibinfo  {journal} {Phys. Rev. Lett.}\ }\textbf {\bibinfo {volume} {110}},\ \bibinfo {pages} {262002} (\bibinfo {year} {2013})},\ \Eprint {https://arxiv.org/abs/1305.1539} {arXiv:1305.1539 [hep-ph]} \BibitemShut {NoStop}%
\bibitem [{\citenamefont {Radyushkin}(2017)}]{Radyushkin:2016hsy}%
  \BibitemOpen
  \bibfield  {author} {\bibinfo {author} {\bibfnamefont {A.}~\bibnamefont {Radyushkin}},\ }\bibfield  {title} {\bibinfo {title} {{Nonperturbative Evolution of Parton Quasi-Distributions}},\ }\href {https://doi.org/10.1016/j.physletb.2017.02.019} {\bibfield  {journal} {\bibinfo  {journal} {Phys. Lett. B}\ }\textbf {\bibinfo {volume} {767}},\ \bibinfo {pages} {314} (\bibinfo {year} {2017})},\ \Eprint {https://arxiv.org/abs/1612.05170} {arXiv:1612.05170 [hep-ph]} \BibitemShut {NoStop}%
\bibitem [{\citenamefont {Gao}\ \emph {et~al.}(2021)\citenamefont {Gao}, \citenamefont {Karthik}, \citenamefont {Mukherjee}, \citenamefont {Petreczky}, \citenamefont {Syritsyn},\ and\ \citenamefont {Zhao}}]{Gao:2021hvs}%
  \BibitemOpen
  \bibfield  {author} {\bibinfo {author} {\bibfnamefont {X.}~\bibnamefont {Gao}}, \bibinfo {author} {\bibfnamefont {N.}~\bibnamefont {Karthik}}, \bibinfo {author} {\bibfnamefont {S.}~\bibnamefont {Mukherjee}}, \bibinfo {author} {\bibfnamefont {P.}~\bibnamefont {Petreczky}}, \bibinfo {author} {\bibfnamefont {S.}~\bibnamefont {Syritsyn}},\ and\ \bibinfo {author} {\bibfnamefont {Y.}~\bibnamefont {Zhao}},\ }\bibfield  {title} {\bibinfo {title} {{Towards studying the structural differences between the pion and its radial excitation}},\ }\href {https://doi.org/10.1103/PhysRevD.103.094510} {\bibfield  {journal} {\bibinfo  {journal} {Phys. Rev. D}\ }\textbf {\bibinfo {volume} {103}},\ \bibinfo {pages} {094510} (\bibinfo {year} {2021})},\ \Eprint {https://arxiv.org/abs/2101.11632} {arXiv:2101.11632 [hep-lat]} \BibitemShut {NoStop}%
\bibitem [{\citenamefont {Holl}\ \emph {et~al.}(2004)\citenamefont {Holl}, \citenamefont {Krassnigg},\ and\ \citenamefont {Roberts}}]{Holl:2004fr}%
  \BibitemOpen
  \bibfield  {author} {\bibinfo {author} {\bibfnamefont {A.}~\bibnamefont {Holl}}, \bibinfo {author} {\bibfnamefont {A.}~\bibnamefont {Krassnigg}},\ and\ \bibinfo {author} {\bibfnamefont {C.~D.}\ \bibnamefont {Roberts}},\ }\bibfield  {title} {\bibinfo {title} {{Pseudoscalar meson radial excitations}},\ }\href {https://doi.org/10.1103/PhysRevC.70.042203} {\bibfield  {journal} {\bibinfo  {journal} {Phys. Rev. C}\ }\textbf {\bibinfo {volume} {70}},\ \bibinfo {pages} {042203} (\bibinfo {year} {2004})},\ \Eprint {https://arxiv.org/abs/nucl-th/0406030} {arXiv:nucl-th/0406030} \BibitemShut {NoStop}%
\bibitem [{\citenamefont {Qin}\ \emph {et~al.}(2012)\citenamefont {Qin}, \citenamefont {Chang}, \citenamefont {Liu}, \citenamefont {Roberts},\ and\ \citenamefont {Wilson}}]{Qin:2011xq}%
  \BibitemOpen
  \bibfield  {author} {\bibinfo {author} {\bibfnamefont {S.-x.}\ \bibnamefont {Qin}}, \bibinfo {author} {\bibfnamefont {L.}~\bibnamefont {Chang}}, \bibinfo {author} {\bibfnamefont {Y.-x.}\ \bibnamefont {Liu}}, \bibinfo {author} {\bibfnamefont {C.~D.}\ \bibnamefont {Roberts}},\ and\ \bibinfo {author} {\bibfnamefont {D.~J.}\ \bibnamefont {Wilson}},\ }\bibfield  {title} {\bibinfo {title} {{Investigation of rainbow-ladder truncation for excited and exotic mesons}},\ }\href {https://doi.org/10.1103/PhysRevC.85.035202} {\bibfield  {journal} {\bibinfo  {journal} {Phys. Rev. C}\ }\textbf {\bibinfo {volume} {85}},\ \bibinfo {pages} {035202} (\bibinfo {year} {2012})},\ \Eprint {https://arxiv.org/abs/1109.3459} {arXiv:1109.3459 [nucl-th]} \BibitemShut {NoStop}%
\bibitem [{\citenamefont {Chang}\ \emph {et~al.}(2014)\citenamefont {Chang}, \citenamefont {Mezrag}, \citenamefont {Moutarde}, \citenamefont {Roberts}, \citenamefont {Rodr\'\i{}guez-Quintero},\ and\ \citenamefont {Tandy}}]{Chang:2014lva}%
  \BibitemOpen
  \bibfield  {author} {\bibinfo {author} {\bibfnamefont {L.}~\bibnamefont {Chang}}, \bibinfo {author} {\bibfnamefont {C.}~\bibnamefont {Mezrag}}, \bibinfo {author} {\bibfnamefont {H.}~\bibnamefont {Moutarde}}, \bibinfo {author} {\bibfnamefont {C.~D.}\ \bibnamefont {Roberts}}, \bibinfo {author} {\bibfnamefont {J.}~\bibnamefont {Rodr\'\i{}guez-Quintero}},\ and\ \bibinfo {author} {\bibfnamefont {P.~C.}\ \bibnamefont {Tandy}},\ }\bibfield  {title} {\bibinfo {title} {{Basic features of the pion valence-quark distribution function}},\ }\href {https://doi.org/10.1016/j.physletb.2014.08.009} {\bibfield  {journal} {\bibinfo  {journal} {Phys. Lett. B}\ }\textbf {\bibinfo {volume} {737}},\ \bibinfo {pages} {23} (\bibinfo {year} {2014})},\ \Eprint {https://arxiv.org/abs/1406.5450} {arXiv:1406.5450 [nucl-th]} \BibitemShut {NoStop}%
\bibitem [{\citenamefont {Chang}(2016)}]{Chang:2016vkv}%
  \BibitemOpen
  \bibfield  {author} {\bibinfo {author} {\bibfnamefont {L.}~\bibnamefont {Chang}},\ }\bibfield  {title} {\bibinfo {title} {{A perspective on Dyson-Schwinger equation: toy model of Pion}},\ }\href {https://doi.org/10.1051/epjconf/201611305001} {\bibfield  {journal} {\bibinfo  {journal} {EPJ Web Conf.}\ }\textbf {\bibinfo {volume} {113}},\ \bibinfo {pages} {05001} (\bibinfo {year} {2016})}\BibitemShut {NoStop}%
\bibitem [{\citenamefont {Xu}\ \emph {et~al.}(2018)\citenamefont {Xu}, \citenamefont {Chang}, \citenamefont {Roberts},\ and\ \citenamefont {Zong}}]{Xu:2018eii}%
  \BibitemOpen
  \bibfield  {author} {\bibinfo {author} {\bibfnamefont {S.-S.}\ \bibnamefont {Xu}}, \bibinfo {author} {\bibfnamefont {L.}~\bibnamefont {Chang}}, \bibinfo {author} {\bibfnamefont {C.~D.}\ \bibnamefont {Roberts}},\ and\ \bibinfo {author} {\bibfnamefont {H.-S.}\ \bibnamefont {Zong}},\ }\bibfield  {title} {\bibinfo {title} {{Pion and kaon valence-quark parton quasidistributions}},\ }\href {https://doi.org/10.1103/PhysRevD.97.094014} {\bibfield  {journal} {\bibinfo  {journal} {Phys. Rev. D}\ }\textbf {\bibinfo {volume} {97}},\ \bibinfo {pages} {094014} (\bibinfo {year} {2018})},\ \Eprint {https://arxiv.org/abs/1802.09552} {arXiv:1802.09552 [nucl-th]} \BibitemShut {NoStop}%
\bibitem [{\citenamefont {Albino}\ \emph {et~al.}(2022)\citenamefont {Albino}, \citenamefont {Higuera-Angulo}, \citenamefont {Raya},\ and\ \citenamefont {Bashir}}]{Albino:2022gzs}%
  \BibitemOpen
  \bibfield  {author} {\bibinfo {author} {\bibfnamefont {L.}~\bibnamefont {Albino}}, \bibinfo {author} {\bibfnamefont {I.~M.}\ \bibnamefont {Higuera-Angulo}}, \bibinfo {author} {\bibfnamefont {K.}~\bibnamefont {Raya}},\ and\ \bibinfo {author} {\bibfnamefont {A.}~\bibnamefont {Bashir}},\ }\bibfield  {title} {\bibinfo {title} {{Pseudoscalar mesons: Light front wave functions, GPDs, and PDFs}},\ }\href {https://doi.org/10.1103/PhysRevD.106.034003} {\bibfield  {journal} {\bibinfo  {journal} {Phys. Rev. D}\ }\textbf {\bibinfo {volume} {106}},\ \bibinfo {pages} {034003} (\bibinfo {year} {2022})},\ \Eprint {https://arxiv.org/abs/2207.06550} {arXiv:2207.06550 [hep-ph]} \BibitemShut {NoStop}%
\bibitem [{\citenamefont {Nakanishi}(1963)}]{Nakanishi:1963zz}%
  \BibitemOpen
  \bibfield  {author} {\bibinfo {author} {\bibfnamefont {N.}~\bibnamefont {Nakanishi}},\ }\bibfield  {title} {\bibinfo {title} {{Partial-Wave Bethe-Salpeter Equation}},\ }\href {https://doi.org/10.1103/PhysRev.130.1230} {\bibfield  {journal} {\bibinfo  {journal} {Phys. Rev.}\ }\textbf {\bibinfo {volume} {130}},\ \bibinfo {pages} {1230} (\bibinfo {year} {1963})}\BibitemShut {NoStop}%
\bibitem [{\citenamefont {Hobbs}\ \emph {et~al.}(2017)\citenamefont {Hobbs}, \citenamefont {Alberg},\ and\ \citenamefont {Miller}}]{Hobbs:2016xfz}%
  \BibitemOpen
  \bibfield  {author} {\bibinfo {author} {\bibfnamefont {T.~J.}\ \bibnamefont {Hobbs}}, \bibinfo {author} {\bibfnamefont {M.}~\bibnamefont {Alberg}},\ and\ \bibinfo {author} {\bibfnamefont {G.~A.}\ \bibnamefont {Miller}},\ }\bibfield  {title} {\bibinfo {title} {{Euclidean bridge to the relativistic constituent quark model}},\ }\href {https://doi.org/10.1103/PhysRevC.95.035205} {\bibfield  {journal} {\bibinfo  {journal} {Phys. Rev. C}\ }\textbf {\bibinfo {volume} {95}},\ \bibinfo {pages} {035205} (\bibinfo {year} {2017})},\ \Eprint {https://arxiv.org/abs/1608.07319} {arXiv:1608.07319 [nucl-th]} \BibitemShut {NoStop}%
\bibitem [{\citenamefont {Navas}\ \emph {et~al.}(2024)\citenamefont {Navas} \emph {et~al.}}]{ParticleDataGroup:2024cfk}%
  \BibitemOpen
  \bibfield  {author} {\bibinfo {author} {\bibfnamefont {S.}~\bibnamefont {Navas}} \emph {et~al.} (\bibinfo {collaboration} {Particle Data Group}),\ }\bibfield  {title} {\bibinfo {title} {{Review of particle physics}},\ }\href {https://doi.org/10.1103/PhysRevD.110.030001} {\bibfield  {journal} {\bibinfo  {journal} {Phys. Rev. D}\ }\textbf {\bibinfo {volume} {110}},\ \bibinfo {pages} {030001} (\bibinfo {year} {2024})}\BibitemShut {NoStop}%
\bibitem [{\citenamefont {Li}\ \emph {et~al.}(2016)\citenamefont {Li}, \citenamefont {Chang}, \citenamefont {Gao}, \citenamefont {Roberts}, \citenamefont {Schmidt},\ and\ \citenamefont {Zong}}]{Li:2016dzv}%
  \BibitemOpen
  \bibfield  {author} {\bibinfo {author} {\bibfnamefont {B.~L.}\ \bibnamefont {Li}}, \bibinfo {author} {\bibfnamefont {L.}~\bibnamefont {Chang}}, \bibinfo {author} {\bibfnamefont {F.}~\bibnamefont {Gao}}, \bibinfo {author} {\bibfnamefont {C.~D.}\ \bibnamefont {Roberts}}, \bibinfo {author} {\bibfnamefont {S.~M.}\ \bibnamefont {Schmidt}},\ and\ \bibinfo {author} {\bibfnamefont {H.~S.}\ \bibnamefont {Zong}},\ }\bibfield  {title} {\bibinfo {title} {{Distribution amplitudes of radially-excited $\pi$ and K mesons}},\ }\href {https://doi.org/10.1103/PhysRevD.93.114033} {\bibfield  {journal} {\bibinfo  {journal} {Phys. Rev. D}\ }\textbf {\bibinfo {volume} {93}},\ \bibinfo {pages} {114033} (\bibinfo {year} {2016})},\ \Eprint {https://arxiv.org/abs/1604.07415} {arXiv:1604.07415 [nucl-th]} \BibitemShut {NoStop}%
\bibitem [{\citenamefont {Li}\ and\ \citenamefont {Vary}(2022)}]{Li:2021jqb}%
  \BibitemOpen
  \bibfield  {author} {\bibinfo {author} {\bibfnamefont {Y.}~\bibnamefont {Li}}\ and\ \bibinfo {author} {\bibfnamefont {J.~P.}\ \bibnamefont {Vary}},\ }\bibfield  {title} {\bibinfo {title} {{Light-front holography with chiral symmetry breaking}},\ }\href {https://doi.org/10.1016/j.physletb.2021.136860} {\bibfield  {journal} {\bibinfo  {journal} {Phys. Lett. B}\ }\textbf {\bibinfo {volume} {825}},\ \bibinfo {pages} {136860} (\bibinfo {year} {2022})},\ \Eprint {https://arxiv.org/abs/2103.09993} {arXiv:2103.09993 [hep-ph]} \BibitemShut {NoStop}%
\bibitem [{\citenamefont {Hoferichter}\ and\ \citenamefont {Stoffer}(2020)}]{Hoferichter:2020lap}%
  \BibitemOpen
  \bibfield  {author} {\bibinfo {author} {\bibfnamefont {M.}~\bibnamefont {Hoferichter}}\ and\ \bibinfo {author} {\bibfnamefont {P.}~\bibnamefont {Stoffer}},\ }\bibfield  {title} {\bibinfo {title} {{Asymptotic behavior of meson transition form factors}},\ }\href {https://doi.org/10.1007/JHEP05(2020)159} {\bibfield  {journal} {\bibinfo  {journal} {JHEP}\ }\textbf {\bibinfo {volume} {05}},\ \bibinfo {pages} {159}},\ \Eprint {https://arxiv.org/abs/2004.06127} {arXiv:2004.06127 [hep-ph]} \BibitemShut {NoStop}%
\bibitem [{\citenamefont {Sultan}\ \emph {et~al.}(2024)\citenamefont {Sultan}, \citenamefont {Kang}, \citenamefont {Bashir},\ and\ \citenamefont {Chang}}]{Sultan:2024mva}%
  \BibitemOpen
  \bibfield  {author} {\bibinfo {author} {\bibfnamefont {M.~A.}\ \bibnamefont {Sultan}}, \bibinfo {author} {\bibfnamefont {J.}~\bibnamefont {Kang}}, \bibinfo {author} {\bibfnamefont {A.}~\bibnamefont {Bashir}},\ and\ \bibinfo {author} {\bibfnamefont {L.}~\bibnamefont {Chang}},\ }\bibfield  {title} {\bibinfo {title} {{Neutral pion to two-photons transition form factor revisited}},\ }\href {https://doi.org/10.1103/PhysRevD.110.114047} {\bibfield  {journal} {\bibinfo  {journal} {Phys. Rev. D}\ }\textbf {\bibinfo {volume} {110}},\ \bibinfo {pages} {114047} (\bibinfo {year} {2024})},\ \Eprint {https://arxiv.org/abs/2409.09595} {arXiv:2409.09595 [hep-ph]} \BibitemShut {NoStop}%
\bibitem [{\citenamefont {Cao}(2021)}]{Cao:2021ddi}%
  \BibitemOpen
  \bibfield  {author} {\bibinfo {author} {\bibfnamefont {F.-G.}\ \bibnamefont {Cao}},\ }\bibfield  {title} {\bibinfo {title} {{Radial excitation of light mesons and the \ensuremath{\gamma}\ensuremath{\gamma}*\textrightarrow{}\ensuremath{\pi}0 transition form factor}},\ }\href {https://doi.org/10.1103/PhysRevD.104.054025} {\bibfield  {journal} {\bibinfo  {journal} {Phys. Rev. D}\ }\textbf {\bibinfo {volume} {104}},\ \bibinfo {pages} {054025} (\bibinfo {year} {2021})},\ \Eprint {https://arxiv.org/abs/2109.02856} {arXiv:2109.02856 [hep-ph]} \BibitemShut {NoStop}%
\bibitem [{\citenamefont {Holl}\ \emph {et~al.}(2005)\citenamefont {Holl}, \citenamefont {Krassnigg}, \citenamefont {Maris}, \citenamefont {Roberts},\ and\ \citenamefont {Wright}}]{Holl:2005vu}%
  \BibitemOpen
  \bibfield  {author} {\bibinfo {author} {\bibfnamefont {A.}~\bibnamefont {Holl}}, \bibinfo {author} {\bibfnamefont {A.}~\bibnamefont {Krassnigg}}, \bibinfo {author} {\bibfnamefont {P.}~\bibnamefont {Maris}}, \bibinfo {author} {\bibfnamefont {C.~D.}\ \bibnamefont {Roberts}},\ and\ \bibinfo {author} {\bibfnamefont {S.~V.}\ \bibnamefont {Wright}},\ }\bibfield  {title} {\bibinfo {title} {{Electromagnetic properties of ground and excited state pseudoscalar mesons}},\ }\href {https://doi.org/10.1103/PhysRevC.71.065204} {\bibfield  {journal} {\bibinfo  {journal} {Phys. Rev. C}\ }\textbf {\bibinfo {volume} {71}},\ \bibinfo {pages} {065204} (\bibinfo {year} {2005})},\ \Eprint {https://arxiv.org/abs/nucl-th/0503043} {arXiv:nucl-th/0503043} \BibitemShut {NoStop}%
\bibitem [{\citenamefont {Carmignotto}\ \emph {et~al.}(2014)\citenamefont {Carmignotto}, \citenamefont {Horn},\ and\ \citenamefont {Miller}}]{Carmignotto:2014rqa}%
  \BibitemOpen
  \bibfield  {author} {\bibinfo {author} {\bibfnamefont {M.}~\bibnamefont {Carmignotto}}, \bibinfo {author} {\bibfnamefont {T.}~\bibnamefont {Horn}},\ and\ \bibinfo {author} {\bibfnamefont {G.~A.}\ \bibnamefont {Miller}},\ }\bibfield  {title} {\bibinfo {title} {{Pion transverse charge density and the edge of hadrons}},\ }\href {https://doi.org/10.1103/PhysRevC.90.025211} {\bibfield  {journal} {\bibinfo  {journal} {Phys. Rev. C}\ }\textbf {\bibinfo {volume} {90}},\ \bibinfo {pages} {025211} (\bibinfo {year} {2014})},\ \Eprint {https://arxiv.org/abs/1404.1539} {arXiv:1404.1539 [nucl-ex]} \BibitemShut {NoStop}%
\end{thebibliography}%
\end{document}